\documentclass[prl,amssymb,twocolumn,tightenlines,showpacs,superscriptaddress]{revtex4}

\usepackage{graphicx}

\begin{document}

\title{Quantum entanglement of a large number of photons}

\author{H.S. Eisenberg}
\affiliation{Department of Physics, University of California, Santa Barbara, California 93106, USA}

\author{G. Khoury}
\affiliation{Department of Physics, University of California, Santa Barbara, California 93106, USA}

\author{G. Durkin}
\affiliation{Department of Physics, University of California, Santa Barbara, California 93106, USA}
\affiliation{Centre for Quantum Computation, University of Oxford, Oxford OX1 3PU, United Kingdom}

\author{C. Simon}
\affiliation{Laboratoire de Spectrom\'{e}trie Physique, CNRS - Universit\'{e} Grenoble I, St.\ Martin d'H\`{e}res, France}

\author{D. Bouwmeester}
\affiliation{Department of Physics, University of California, Santa Barbara, California 93106, USA}



\begin{abstract}
A bipartite multiphoton entangled state is created through
stimulated parametric down-conversion of strong laser pulses
in a nonlinear crystal. It is shown how detectors that do not
resolve photon number can be used to analyze such multiphoton
states. Entanglement of up to 12 photons is detected using
both the positivity of the partially transposed density matrix
and a newly derived criteria. Furthermore, evidence is
provided for entanglement of up to 100 photons. The
multi-particle quantum state is such that even in the case of
an overall photon collection and detection efficiency as low
as a few percent, entanglement remains and can be detected.
\end{abstract}

\maketitle

In recent years, small numbers of entangled particles have
been used for testing quantum mechanics and for implementing
various quantum information protocols\cite{Book}. However,
other tests probing the validity of quantum decoherence models
\cite{Auffeves}, and additional quantum information protocols
will require entangled states of large numbers of particles.
Bipartite multiphoton states, the subject of this paper, can
be used to test foundations of quantum
theory\cite{Drummond,Howell}, and for quantum
cryptography\cite{Durkin}. Furthermore, it has been shown that
phase sensitive measurements\cite{Holland} and quantum
photolithography\cite{Boto} can exceed classical boundaries
imposed by the wavelength of light, by using multiple
entangled photons\cite{Mitchell}.

In this Letter we demonstrate the generation of a bipartite
entangled state of many photons in two spatial modes, as
produced by stimulated parametric down-conversion
(PDC)\cite{Kwiat,Kok,Lamas-Linares,Simon}. The Hamiltonian for
the generation of polarization entangled photons \cite{Kwiat}
is given by
\begin{equation}\label{Hamiltonian}
H=i\kappa(a_h^\dag b_v^\dag-a_v^\dag b_h^\dag)+h.c.\,
\end{equation}
Horizontally ($h$) and vertically ($v$) polarized photons occupy two
spatial modes (\textit{a} and \textit{b}); $\kappa$ is a coupling
constant that depends on the nonlinearity of the
crystal and the intensity of the pump pulse. The resulting photon
state is given by\cite{Kok}
\begin{eqnarray}\label{Psi}
|\psi\rangle&=&\frac{1}{\cosh^{2}\tau}\sum_{n=0}^\infty\sqrt{n+1}\tanh^n\tau|\psi^-_n\rangle\,,\\
\nonumber
|\psi^-_n\rangle&=&\frac{1}{\sqrt{n+1}}\sum_{m=0}^n(\textendash
1)^m|n\textendash
m\rangle_{a_h}|m\rangle_{a_v}|m\rangle_{b_h}|n\textendash
m\rangle_{b_v}\,,
\end{eqnarray}
where, for example, $|m\rangle_{a_v}$ represents $m$
vertically polarized photons in mode $a$. The interaction
parameter $\tau$ depends linearly on the crystal length and on
$\kappa$. The state $|\psi\rangle$ is a superposition of the
states $|\psi^-_n\rangle$ of $n$ indistinguishable photon
pairs. Each $|\psi^-_n\rangle$ is an analog of a singlet state
of two spin-$n/2$ particles, thus $|\psi\rangle$ is invariant
under joint rotations of the polarization bases of both modes.
The average photon pair number is $\langle
n\rangle=2\sinh^2\tau$. Although photons are created in pairs,
the resulting state cannot be factorized into individual
pairs. As a result of the stimulated emission process, the
pairs are indistinguishable such that they form a single
multiphoton entangled state. Previous PDC experiments have
been restricted to $\tau < 0.1$, resulting in the detection of
at most 4 to 5 photons in an entangled
state\cite{Lamas-Linares,Eibl,Pan}. This work addresses the
region of $\tau > 1$, where entangled states of large numbers
of photons can be generated.

\begin{figure}[tbp]
\includegraphics[angle=-90,width=76mm]{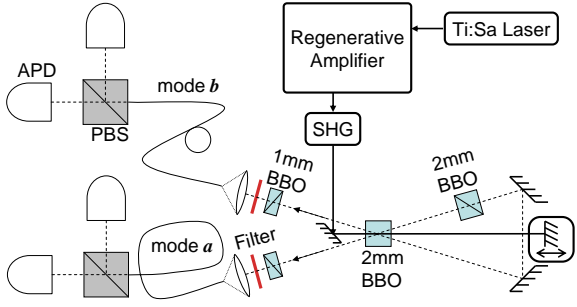}
\caption{\label{fig0}The experimental setup. The pump pulses pass
twice through the nonlinear BBO crystal, with a controlled delay
between the passes. Down-converted photons from the first pass are
re-injected into the crystal together with the pump second
pass\cite{Lamas-Linares}. The configuration can be switched to
single-pass by blocking the path of the first down-conversion.
Additional BBO crystals are inserted in order to compensate for
temporal walk-off. The two spatial modes are collected into
single-mode fibers through narrow bandpass filters. Photons in the
different modes are detected by four APDs.}
\end{figure}

Our setup is switchable between single-pass and
double-pass\cite{Lamas-Linares} of a pump pulse through a BBO
nonlinear crystal (see Fig. \ref{fig0}). The pump is a
frequency-doubled amplified Ti:sapphire laser, giving 200\,fs
pulses with 5\,$\mu$J per pulse. Two down-converted modes $a$
and $b$ are selected by coupling into single-mode fibers
through 5\,nm narrow bandpass filters, justifying the use of
the Hamiltonian of Eq. \ref{Hamiltonian}. For each spatial
mode, two orthogonal polarizations are separated by a fiber
polarization beam splitter and detected by silicon avalanche
photodiodes (APD). Single-photon count rates and coincidence
rates were recorded as functions of the pump power in three
polarization bases: horizontal/vertical linear ($hv$),
plus/minus $45^\circ$ linear ($pm$) and right/left circular
($rl$).

To characterize states of multiple photons, it is desirable to
have detectors that can resolve photon number. Although such
detectors exist\cite{Turner,Kim,Gol'tsman,Waks,Achilles}, the
photon number resolution is always limited by losses.
Therefore, to determine the actual multiphoton state produced
at the source, it is necessary to perform a probability
analysis of the experimental data, based on the physics of the
detection scheme.

We use APDs that give no direct information about the number
of detected photons. Nevertheless, the probability to obtain a
signal depends on the photon number\cite{Mogilevtsev,Resch}.
For a total collection efficiency $\eta$ (a combination of the
APD detectors and the optical coupling efficiencies), the
triggering probability given an $m$-photon state is
\begin{equation}\label{P1General}
P=1-P_0=1-(1-\eta)^m\,,
\end{equation}
where $P_0$ is the probability of not detecting a photon.

For the state of Eq. \ref{Psi}, the detection probabilities for a
single spatial mode and coincidences between any two modes as
functions of $\tau$ and $\eta$ were derived. For example, the
probability per pulse to trigger a single detector is
\begin{eqnarray}\label{P1}
P&=&\frac{1}{\cosh^{4}\tau}\sum_{n=1}^\infty\tanh^{2n}\tau\sum_{m=1}^n(1-(1-\eta)^m)\\
\nonumber &=&\frac{\eta\tanh^2\tau}{1-(1-\eta)\tanh^2\tau}\,.
\end{eqnarray}
We measured probabilities as functions of the pump intensity $I$ up
to a maximal intensity $I_{max}$, and fitted the results with the
collection efficiencies $\eta_i$ of the four modes $a_h$, $a_v$,
$b_h$ and $b_v$, and the maximum interaction parameter $\tau_{max}$
defined as $\tau_{max}=\tau\sqrt{I_{max}/I}$.

\begin{figure}[tbp]
\includegraphics[angle=0,width=76mm]{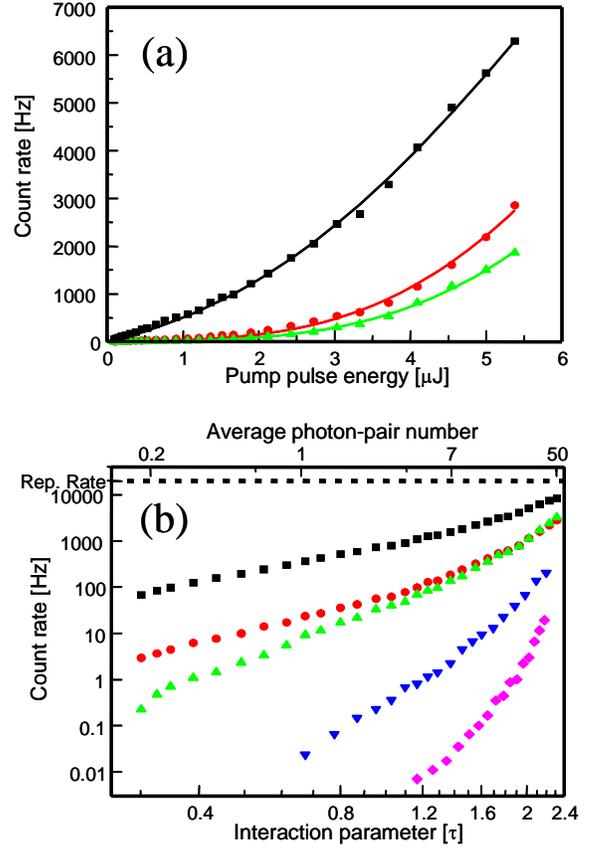}
\caption{\label{fig1}(\textbf{a}) Experimentally observed
single photon count rates of the $a_h$ mode
(\textbf{squares}), one-pair coincidence rates between $a_h$
and $b_v$ (\textbf{circles}) and two-pair coincidence rates
between $a_h$ and $b_h$ (\textbf{triangles}) as a function of
the pump pulse energy. The one-pair coincidences can arise
from one or more pairs, while the two-pair coincidences can
arise from two or more. Fits are included as solid lines.
(\textbf{b}) The $\tau$ dependance of the results of Fig.
\ref{fig1}a and coincidence events that can be generated by at
least three pairs (\textbf{inverted triangles}) and four pairs
(\textbf{diamonds}) of photons.}
\end{figure}

Single-photon counts of one polarization mode and coincidence
counts are presented in Fig. \ref{fig1}a. This data is from a
single-pass configuration. All the results in various
polarization bases were successfully fitted with the same
parameters $\tau_{max}=2.30\pm0.05$ and $\eta=1.9\pm0.2\%$ for
all four modes, strongly supporting the model of Eq.
(\ref{Hamiltonian}). The stimulated emission enabled the
direct observation of coincidences that can only occur from
events of at least three or four pairs. By collecting only one
polarization and splitting the photons into two detectors with
a beam splitter, we counted coincidence events of the form
$a_h$-$a_h$-$b_h$ and of the form $a_h$-$a_h$-$b_h$-$b_h$ that
originated from three (or more) and four (or more)
photon-pairs, respectively. Figure \ref{fig1}b combines all
the measurements as a function of $\tau$. The larger the
number of relevant pairs for an event, the steeper is its
graph, as expected from a multiphoton stimulated process. The
slopes for $\tau < 1$ of the single photon counts and the
one-pair coincidence are parallel and linear with pump pulse
energy, as both result from one-pair events. All the graphs
should saturate for large enough $\tau$ at the repetition rate
of 20\,kHz. The maximum interaction parameter achieved
corresponds to 100 photons per pulse on average.

We have shown a stimulated emission process in which many
photons are created in a way consistent with the state of Eq.
\ref{Psi}. This does not yet prove the specific quantum
correlations between the photons described by that state. We
will now present two criteria that prove the presence of
entanglement. Our approach is to use a low overall detection
efficiency such that in the relevant parameter regime we
detect at most two photons. We show that entanglement is still
present in this situation. This proves entanglement not only
of the detected photons, but also of the initial state before
most of the photons were lost, as it is impossible to form an
entangled state by applying a local operation to unentangled
photons.

\begin{figure}[tbp]
\includegraphics[angle=-90,width=76mm]{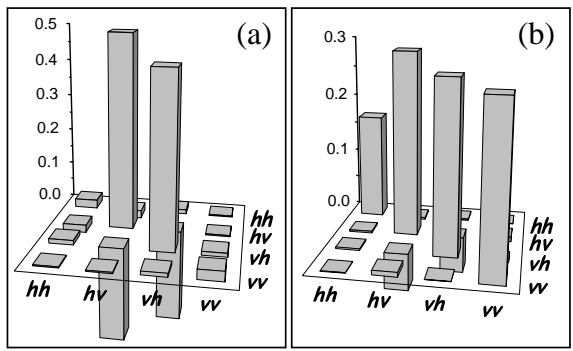}
\caption{\label{fig3}Measured density matrices in the (1,1)
subspace for $\tau=0.2$ (\textbf{a}) and $\tau=1.85$
(\textbf{b}). Only the real part of $\rho$ is shown as its
imaginary part is negligible.}
\end{figure}

The Positivity of the Partially Transposed (PPT) density
matrix is a separability (non-entanglement) criterion for
bipartite systems, such as the PDC system studied
here\cite{Peres}. Consider the density matrix $\rho$ in the
subspace where only one photon after losses is detected in
each of the spatial modes. This restriction will be justified
below. The total probability for such a detection is
$P_{(1,1)}=P_{hv}+P_{vh}+P_{hh}+P_{vv}$, where $P_{hv}$ is the
probability to detect a coincidence between $a_h$ and $b_v$,
etc. By only considering events from this (1,1) subspace,
probabilities can be normalized as $p_{ij}=P_{ij}/P_{(1,1)}$.
We use $|hh\rangle$, $|hv\rangle$, $|vh\rangle$ and
$|vv\rangle$ as the basis states for $\rho$. We also define
the single-pair visibility $V$ for different polarization
detection bases. For example, in the $hv$ basis for mode $a$
and the $pm$ basis for mode $b$
\begin{equation}\label{Vdef}
V_{hv,pm}=\frac{P_{hm}+P_{vp}-P_{hp}-P_{vm}}{P_{hm}+P_{vp}+P_{hp}+P_{vm}}\,.
\end{equation}

The elements of $\rho$ can be readily obtained from
combinations of visibilities, a process known as state
tomography\cite{James}. Density matrices, as measured for low
($\tau=0.2$) and high ($\tau=1.85$) values of $\tau$, are
presented in Fig. \ref{fig3}. The measured density matrices
are consistent with the state of Eq. \ref{Psi}; for low tau
the density matrix approaches the familiar two-photon
$|\psi^-_1 \rangle$ Bell state
($\rho_{hv,hv}=\rho_{vh,vh}=-\rho_{hv,vh}=-\rho_{vh,hv}=1/2$),
for high tau $hh$ and $vv$ coincidences are detected
($\rho_{hh,hh}$ and $\rho_{vv,vv}$ are no longer small) as a
result of multiple photon pairs before detection losses.
Considering these dominant 6 terms, the partially-transposed
matrix $\rho^{PT}$ will only have positive eigenvalues if
\begin{eqnarray}\label{cg2}
C_1=\frac{16\cdot
p_{hh}p_{vv}}{(V_{pm,pm}+V_{rl,rl})^2+(V_{pm,rl}-V_{rl,pm})^2}>1\,.
\end{eqnarray}

\begin{figure}[tbp]
\includegraphics[angle=0,width=76mm]{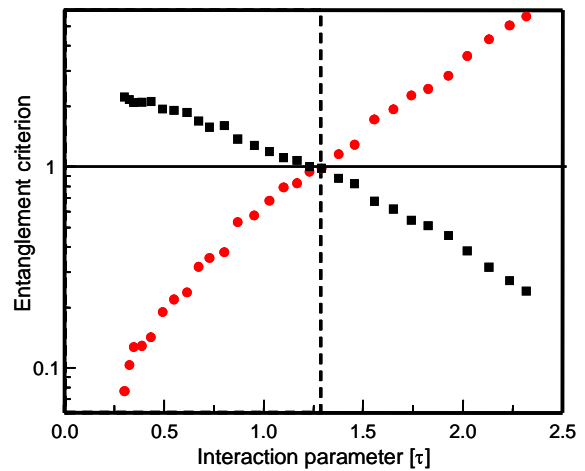}
\caption{\label{fig4}The experimentally measured entanglement
criteria. To detect entanglement the PPT criterion $C_1$
(\textbf{circles}) must be smaller than one, while the
visibility/spin-correlation criterion $C_2$ (\textbf{squares})
must be larger than one. Both criteria detect entanglement up
to an interaction of $\tau=1.3$ (\textbf{dashed line}),
corresponding to a state with 12 photons on average.}
\end{figure}

The violation of the above separability criterion proves
entanglement. For example, for a pure $|\psi_1^-\rangle$ state
$C_1$ is zero. The circles in Fig. \ref{fig4} show the
measured $C_1$ as a function of $\tau$. Entanglement is
detected up to an interaction of $\tau=1.3$, corresponding to
6 indistinguishable photon-pairs on average. For this $\tau$
value and detection efficiency of 2\%, the ratio of the
probabilities for a detection from a higher photon number
subspace and from a (1,1) subspace is about $0.06$. This low
ratio justifies considering only (1,1) subspace events for the
density matrix.

We now derive a second separability criterion tailored to
detect the type of entanglement created by PDC. The visibility
(Eq. \ref{Vdef}) can be rewritten as the spin anti-correlation
between the two spatial modes\cite{Simon}:
\begin{equation}\label{VisAC}
V_{hv,hv}=p_{hv}+p_{vh}-p_{hh}-p_{vv}=-\left<\sigma^a_z\otimes\sigma^b_z\right>\,,
\end{equation}
where $\sigma_i$ are the Pauli matrices. The total spin
correlation is:
\begin{eqnarray}\label{TotalCorr}
\left<\overrightarrow{\sigma}^a\cdot\overrightarrow{\sigma}^b\right>&=&\left<\sigma^a_x\otimes\sigma^b_x\right>+\left<\sigma^a_y\otimes\sigma^b_y\right>+\left<\sigma^a_z\otimes\sigma^b_z\right>\\
\nonumber &=&-(V_{pm,pm}+V_{rl,rl}+V_{hv,hv})\,.
\end{eqnarray}
A product state, is maximally correlated (anti-correlated)
when the two spins are parallel (anti-parallel). It is
convenient to rotate the two correlated spins to one of the
principal bases. The correlation in that basis will be $\pm1$,
and zero in the other two bases. A general separable state (a
mixture of product states) can not have a higher correlation.
On the other hand, an entangled state such as the Bell
$|\psi^-_1 \rangle$ state can have all the correlations as -1.
Thus, an upper bound criterion for a separable state in the
(1,1) subspace is
\begin{equation}\label{C1}
C_2=|V_{pm,pm}+V_{rl,rl}+V_{hv,hv}|\leq1\,.
\end{equation}
The measured values of $C_2$ are shown as squares in Fig.
\ref{fig4}. This second criterion detects entanglement also up
to $\tau=1.3$, or up to about 6 photon pairs. Although based
on different arguments, the two criteria detect entanglement
up to the same interaction values. This boundary is mainly set
by the limitations of the APDs and not by the actual
entanglement present in the generated state. One can show
theoretically that some entanglement remains for arbitrarily
high $\tau$ and losses \cite{Simon,QuantPH}.

\begin{figure}[tbp]
\includegraphics[angle=0,width=76mm]{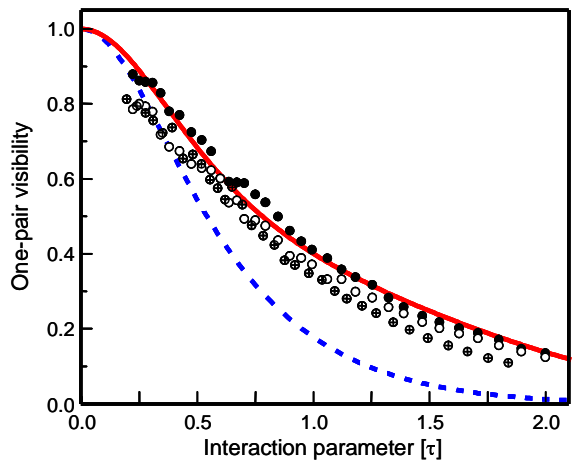}
\caption{\label{fig5}One-pair visibility as a function of the
interaction parameter for the polarization bases $hv$
(\textbf{solid circles}), $pm$ (\textbf{open circles}), and
$rl$ (\textbf{crossed circles}), and for the model fit
(\textbf{solid line}). The prediction for the ansatz state of
distinguishable entangled pairs (\textbf{dashed line})
represents the visibility upper bound for uncorrelated pairs
of photons. The observed results are above this bound,
indicating the indistinguishable nature of the generated
photon-pairs.}
\end{figure}

To provide experimental support for entanglement over the
entire detected interaction range, the one-pair visibilities
in three polarization bases as measured with a double-pass
setup, optimized for collection efficiencies of $\eta=9 \pm
0.7\%$, are shown in Fig. \ref{fig5} and compared to their
theoretical prediction (\textbf{solid line}). The
$\tau$-dependence is approximately the same in the different
polarization bases, consistent with the state rotation
invariance.  For comparison, the visibility for an ansatz
state of \textit{distinguishable} pairs of entangled photons
was also calculated (see \textbf{dashed line} in Fig.
\ref{fig5}). We used the same pair-number distribution as in
PDC, but assumed that the different pairs occupy different
modes. The predicted visibility for this ansatz state is
considerably lower than the PDC visibility curve and the
experimental results.

In conclusion we have demonstrated the generation of a
bipartite state of up to 50 indistinguishable photon pairs.
Entanglement up to 12 photons has been proven while evidence
has been given for entanglement up to 100 photons. We have
shown that it is possible to explore quantum entanglement even
after the state suffered significant losses and even with
detectors that have limited photon number resolution. The
studied multiphoton entangled state is of particular interest
for quantum cryptography, quantum metrology and for tests of
the foundations of quantum mechanics with large-spin systems.

The authors thank M. de Dood and W. Irvine for fruitful discussion.
This research has been supported by NSF no 0304678 and DARPA MDA
972-01-1-0027 grants.

\end{document}